\documentclass[12pt]{article}
\title{Quantum Mechanics as a Principle Theory}
\author{Jeffrey Bub\thanks{Committee on the History and Philosophy of 
Science, University of Maryland, College Park, Maryland 27402, 
U.S.A. (email: jbub@carnap.umd.edu).}}
\date{}

\begin{document}
\maketitle

\begin{abstract} 
I show how quantum mechanics, like the theory of relativity, 
can be understood as a `principle theory' in Einstein's sense, 
and I use this notion to explore the approach to the problem of 
interpretation developed in my book \emph{Interpreting the Quantum 
World}.
\end{abstract}

\bigskip
\bigskip

In an article written for the London \emph{Times} in 1919, Einstein 
presented a remarkably clear and succinct, nontechnical account of the 
theory of relativity. He began by drawing a distinction between 
`constructive' and `principle' theories, and pointed out that the 
theory of relativity should be understood as a principle theory `in 
order to properly grasp its nature.' (I use the translation 
in Einstein (1954), which is different 
from the original translation in the \emph{Times}. See Einstein (1954, 
p. 228).)

Here I apply Einstein's distinction to the problem of interpreting 
quantum mechanics. I show how quantum 
mechanics, too, can be understood as a principle theory, and I 
examine some puzzling features of `entangled states' from this perspective.

\section{Principle and Constructive Theories}

Most theories in physics are constructive theories. As Einstein puts 
it (1954, p. 228),
constructive theories attempt `to build up a picture of the more 
complex phenomena out of the materials of a relatively simple formal 
scheme from which they start out.' His example here is 
the kinetic theory of gases, which `seeks to reduce mechanical, 
thermal, and diffusional processes to movements of molecules---i.e., 
to build them up out of the hypothesis of molecular motion.'

By contrast, principle theories `employ the analytic, not the 
synthetic, method. The elements which form their basis and 
starting-point are not hypothetically constructed but empirically 
discovered ones, general characteristics of natural processes, 
principles that give rise to mathematically formulated criteria which 
the separate processes or the theoretical representations of them have 
to satisfy.' His example is thermodynamics, which 
`seeks by analytical means to deduce necessary conditions, which 
separate events have to satisfy, from the universally experienced fact 
that perpetual motion is impossible.'

This is the picture: A constructive theory, like the kinetic theory 
of gases, begins with certain hypothetical elements, the elementary 
building blocks in terms of which we attempt to construct models of 
more complex processes. So the fundamental theoretical problem for a 
constructive theory is how to synthesize complex processes out of the 
elementary building blocks of the theory, or how to reduce complex 
phenomena to the properties of these elementary building blocks. 

The starting point of a principle theory is a set of empirical laws or 
principles, which may conflict with one another from the standpoint 
of current theory. The fundamental 
theoretical problem is the analysis of these principles, with the aim 
of arriving at certain necessary conditions or constraints on events 
in a theoretical framework that can be seen as underwriting and 
reconciling these empirical principles. We ask: what must the world be 
like---what are the necessary conditions on events---if certain 
empirical laws are to hold.

To show that a theory is a principle theory, we first need to identify 
the empirical principles on which it is based. In the case of special 
relativity, there are two: the principle of relativity and the 
principle of the constancy of the velocity of light. The 
(special) principle of relativity says that the Newtonian principle of 
relativity---the equivalence of inertial observers or inertial 
reference frames for all dynamical phenomena---applies to the whole 
of physics, including electromagnetic phenomena. Drinking a glass of 
wine or watching a movie on a transatlantic flight is no different 
from doing these things on earth, except during turbulence when the 
plane is no longer an inertial system. The principle of the constancy 
of the velocity of light says that the velocity of light in a vacuum 
is independent of the velocity of the observer or the velocity of the 
light source. As Bondi puts it (1964, 1967), velocity doesn't matter, 
and there is 
no overtaking of light by light in empty space.

A consequence of these principles is that there is a difference in 
the Doppler effect for sound and light. If Alice and Bob are two 
inertial observers, and Alice transmits sound signals to Bob with a 
constant time interval $\tau_{t}$ between successive signals, Bob will 
receive these signals with a time interval $\tau_{r}$ between 
successive signals, where $\tau_{r}$ depends not only on their 
relative velocity, but on the 
velocity of each relative to the air, the medium of transmission.  In 
the case of light signals between two inertial observers, there is no 
light medium---no ether---corresponding to the medium of 
transmission for sound. So for light, the ratio of the interval of 
reception to the interval of transmission, $k = \tau_{r}/\tau_{t}$, is constant 
and depends 
only on the relative velocity of the transmitter and receiver. It 
follows immediately that different inertial observers assign different 
time intervals  to the elapsed time between two events, as measured 
by their own clocks, and hence different distances to spatially 
separated events. For an account of relativity in terms of the 
`$k$-calculus,' see Bondi (1964, 1967).

The special theory of relativity modifies classical kinematics to 
provide a framework that incorporates the principle of relativity and 
the principle of the constancy of the velocity of light. Formally, 
the Euclidean space-time geometry of Newtonian mechanics is replaced 
by Minkowskian geometry. The modification of classical kinematics 
requires certain changes in classical dynamics that entail, among 
other consequences, the
equivalence of inertial mass and energy.

\section{From Classical to Quantum Mechanics}

As Bohr (1935, 1961, 1966) saw it, quantum mechanics is a 
`rational generalization' of 
classical mechanics, incorporating the quantum postulate and the correspondence 
principle.
Quantum mechanics as 
a principle theory is the end product of an analysis that begins with 
these principles.

According to classical electromagnetic theory, an electron in periodic 
motion about the nucleus of an atom will emit light in virtue of its 
acceleration. A general (possibly non-circular) orbit can be described in terms 
of a fundamental mode and higher harmonics, analogous to the 
Fourier analysis of the periodic motion of a string. Just as the 
frequencies of the higher harmonics of a vibrating string are integral 
multiples of the fundamental frequency, so the frequencies of the 
light waves emitted 
by an orbiting electron are integral multiples of the fundamental frequency 
of the 
electron's orbital motion. On the classical theory, waves with the frequencies of 
the higher harmonics are emitted simultaneously with waves with the frequency of 
the fundamental mode. 

What we find empirically is a pattern of frequencies in the light 
emitted by a radiating atom that cannot be accounted for by the 
classical theory. On Bohr's theory of the atom, which provides a 
recipe for calculating the observed frequencies $\nu$,  an atom radiates when an 
electron jumps between orbits or `stationary states' 
associated with discrete or quantized 
values of the energy, $E_{n}$, such that:
\[
E_{n} - E_{m} = h\nu
\]

For small values of $n$, there is no relation between the frequencies of 
light emitted by an atom and the frequencies of the harmonics of the 
electron's orbital motion. For large values of $n$, when the classical orbits 
are close together, 
the frequencies of emitted light are all (approximately) integral 
multiples of the fundamental frequency of the $n$'th orbital motion 
of the electron, as in the classical theory, but the higher 
harmonics are not emitted simultaneously with the fundamental mode. 
Either a wave with the fundamental frequency of 
the electron's orbital 
motion in that stationary state is emitted in a transition (corresponding to the 
transition from one stationary state to an adjacent 
stationary state), or a wave with a frequency corresponding to one of the 
higher harmonics is emitted (corresponding to a transition between 
non-adjacent stationary states). This 
is the correspondence principle: For large values of $n$, the 
frequencies for the transitions $n \rightarrow n-1, n \rightarrow 
n-2, \ldots$ 
correspond to the frequencies of the fundamental mode and the 
successively higher harmonics of the Fourier series for the classical 
motion of the electron in the $n$'th stationary state. A similar 
correspondence applies to the intensities and polarizations of the 
light.

Bohr's theory incorporates the quantum postulate for the frequencies 
of the radiation emitted by an atom and 
provides a recipe for calculating transition 
probabilities between stationary states, but without introducing a 
mechanism for the transitions. Tomonaga (1968, pp. 159--60) 
discusses what he calls `the viewpoint of common sense,' that `this procedure 
is to be considered as merely a 
convenient recipe for calculating quantum theoretical quantities such 
as transition probabilities, since we do not know the cause of a 
quantum transition \ldots [and when] \ldots  we arrive at the true theory, 
the first thing 
to be clarified is the mechanism through which only a certain discrete 
set of states can occur in nature and then to understand what 
determines why some atoms jump from $A$ to $B$ at time $t_{B}$ while 
others jump from $A$ to $B'$ at $t_{B'}$, and so on.' He comments
(1968, p. 160; my italics):

\begin{quotation}
It is very natural to anticipate such a future for the quantum theory. 
However, in applying the correspondence principle to various problems, 
a group of physicists, with Bohr as leader, began to think 
differently. Namely, they began to realize that the nature of the 
discontinuities or of the transitions should be sought in the 
correspondence principle itself and that there are no \ldots 
fundamental laws which have no correspondence to the classical theory. 
According to the viewpoint of common sense, a hidden mechanism is to 
exist to make the states discontinuous, and there should be laws of a 
more fundamental nature which describe the course of a transition, 
but this viewpoint should be abandoned. The correspondence principle 
so far described is still too vague in its formulation, but, 
following Bohr, it is anticipated that \emph{the correct laws of the 
quantum world should be obtained not by introducing certain 
additional laws for the transition mechanism but instead by a revised 
form, expressed mathematically in a clear-cut way, of the 
correspondence principle itself.} 
\end{quotation}

Heisenberg, in developing his matrix version of quantum 
mechanics in 1925, asked how the kinematics of classical mechanics 
could be modified so as to yield Bohr's frequency condition for the 
radiation emitted by an atom when an electron jumps between orbits for 
small values of $n$, and the correspondence principle for large values 
of $n$. Beginning with these two `empirically discovered' principles, 
Heisenberg arrived at a theory of motion in 
which the representatives of certain classical 
dynamical variables do not commute. 
This replacement of a commutative algebra of dynamical variables with a 
noncommutative algebra turned out to involve replacing the representation 
of dynamical properties by the subsets of a set---the phase space of 
classical mechanics---with the 
representation of these properties by the subspaces of a vector 
space, the Hilbert space of quantum mechanics. 
That is, it involves the representation of dynamical properties 
by a non-Boolean algebra of a 
certain sort (the lattice of 
subspaces of a suitable Hilbert space) instead of by a Boolean algebra 
(the Boolean algebra of subsets of phase 
space). 
The salient structural feature of the transition from classical to 
quantum mechanics, as von Neumann saw, is the 
replacement of a set-theoretical or Boolean structure for the 
representation of the properties of a mechanical system with a 
projective geometry. This structural change introduces a new element, 
the \emph{angle} between subspaces representing properties, that 
is not present in a set-theoretic representation. The angles are 
related to probabilities---in fact, by Gleason's theorem, to the only 
way probabilities can be defined on a non-Boolean structure of this 
sort. For details see, for example, the discussion in 
the Coda of Bub (1997).

Shortly after the appearance of Heisenberg's matrix mechanics, 
Schr\"{o}\-dinger developed a wave mechanical version of quantum 
mechanics from the idea of wave-particle duality proposed by de 
Broglie and proved the formal equivalence of the two theories. (For 
more details, going beyond the present sketch, see M\"{u}ller (1997).)
Physicists tended to adopt the wave theory as a new way of modelling the 
micro-world and regarded Heisenberg's noncommutative mechanics as a 
formally equivalent version of wave mechanics, without any 
special foundational significance. But the significance of the transition 
from classical to quantum mechanics is quite different from the 
perspective of these two versions of quantum mechanics. 
Schr\"{o}dinger initially proposed his theory as a new constructive 
theory of the processes occurring in atoms, in terms of standing waves 
as the fundamental building blocks. By contrast, Heisenberg's theory is clearly 
formulated as a principle theory in Einstein's sense: `a revised 
form, expressed mathematically in a clear-cut way, of the 
correspondence principle itself.' In the case of 
relativity, the modifications relative to classical mechanics involve 
geometric structure. For quantum mechanics, the modifications involve 
logical structure, in the sense of the possibility structure of 
events: the network of constraints on the possible combinations of properties. 

In the following section, I pose the question of how to make sense of 
the notion of possibility in a non-Boolean world. I suggest that the 
Copenhagen interpretation---at least on Bohr's version---takes quantum 
mechanics as a principle theory, with the `Kantian' twist, based on a 
transcendental argument from the primacy of classical concepts, that 
we only have access to the non-Boolean quantum world through Boolean 
perspectives provided by our classically describably experimental contexts. 
I shall argue that Booleanity is not required to preserve much of our 
commonsense realist intuitions, and that the significance of interpreting quantum 
mechanics as a principle theory is that the possibility structure of 
events in a quantum universe is not fixed, 
as in a classical universe, but changes 
dynamically (just as the significance of the transition from 
classical mechanics to relativistic mechanics is that the geometry of 
our universe is not a fixed Euclidean geometry, but a non-Euclidean 
geometry that changes dynamically with the distribution of mass in the 
universe). In section 4, I show how we can understand quantum 
`entanglement' as a feature of the dynamical evolution of the 
possibity structure of composite systems, and I illustrate this in 
terms of the phenomenon of quantum teleportation.

\section{Quantum Mechanics as a Principle Theory of Logical Structure}

In a classical world characterized by a Boolean possibility 
structure, there are in general many possible truth-value assignments, 
defined by 
2-valued homomorphisms on the algebra, that assign every proposition a 
truth value, either true or false. A classical state can be understood 
as encoding a catalogue of properties of a classical system, specified 
by the true propositions defined by a 2-valued homomorphism. The 
equations of motion in a classical theory describe how these 
properties change over time, that is, how what is actually the case at 
a certain time changes to what is actually the case at a later time. 

A quantum world is characterized by a specific sort of non-Boolean 
possibility structure, represented by the lattice, ${\cal L}$, of 
subspaces of a Hilbert space. This lattice is also a partial Boolean 
algebra, that is, it can be represented as a family of Boolean 
algebras or lattices pasted together in a certain way (with maximum 
and minimum elements identified, and such that, for every pair of 
commuting elements, there exists a Boolean algebra in the family 
containing both elements). There are no 2-valued maps on ${\cal L}$ 
that reduce to 2-valued homomorphisms on each Boolean subalgebra of 
${\cal L}$, except in the case of spin-$\frac{1}{2}$ quantum systems. This 
is the import of the Kochen and Specker (1967) `no go' hidden variable 
theorem. Bell's (1964) `no go' theorem shows that for separated 
spin-$\frac{1}{2}$ systems, $A$ and $B$, the quantum statistics for 
`entangled' quantum states of $A$ and $B$ cannot be recovered from 
measures over 2-valued maps on ${\cal L}_{A+B}$ that are 
2-valued homomorphisms locally on each Boolean subalgebra of ${\cal L}_{A}$ 
and ${\cal L}_{B}$. So the statistics of entangled 
quantum states for spatially separated systems cannot be reduced to 
measures over possible catalogues of properties for each system 
separately, even for spin-$\frac{1}{2}$ systems. 

How, then, do we introduce notions of actuality, possibility, and 
probability on the non-Boolean structure ${\cal L}$? The textbook
position, following Dirac and von Neumann, is to take the quantum 
analogue of the classical state as represented by a unit vector or 
ray in Hilbert space, where the catalogue of properties selected by 
the ray are those properties represented by subspaces containing the 
ray, that is, properties assigned unit probability by the quantum 
state. The corresponding propositions are taken as true of the system in the 
given state. Propositions assigned probability zero by the state are 
taken as false, while other propositions are taken as neither true 
nor false. This position, given the linearity of the equations of 
motion, leads immediately to the measurement problem. 
(See, for example, Bub (1997, Chapter 1).) To avoid 
the measurement problem, Dirac and von Neumann proposed that 
a measurement in quantum mechanics introduces 
a discontinuous and stochastic `collapse' or `jump' of 
the quantum state onto the subspace corresponding to the property 
registered in the measurement, with a probability equal to the probability of 
the property as specified by the state. This move requires that 
measurement processes are somehow distinguished from other processes 
in a quantum world, insofar as they involve the suspension of the 
linear deterministic quantum dynamics in favour of stochastic 
`collapses' or `quantum jumps.'
 
Ever since the 
Solvay conference of October, 1927, most physicists have payed
lip service to the Copenhagen interpretation 
of Bohr and Heisenberg as the `orthodox' interpretation of quantum 
mechanics, but the Copenhagen interpretation differs substantially 
from the interpretation of Dirac and von Neumann. 
On Bohr's version, a quantum `phenomenon' is 
an individual process that occurs under conditions defined by a 
specific, classically describable experimental arrangement, and an 
observable can be said to have a determinate value only in the 
context of an experiment suitable for measuring the observable. The 
experimental arrangements suitable for locating an atomic object in 
space and time, and for a determination of momentum-energy values, 
are mutually exclusive. We can choose to investigate either of these 
`complementary' phenomena at the expense of the other, so there is 
no unique description of the object in terms of determinate 
properties.
 
The Copenhagen interpretation takes quantum mechanics as 
a principle theory rather than a constructive theory. But the 
analysis is qualified by the assumption that determinateness is only 
meaningful in a Boolean context, defined by the possible values of a 
single maximal observable (or complete commuting set of observables) 
associated with the complete specification of a
classically describable experimental arrangement. In effect, the view 
is that we only have access to the non-Boolean quantum world through Boolean 
`windows,' defined by the behaviour of clasically describable 
macrosystems in their r\^{o}le as measuring instruments. 
The underlying idea--which I referred to as `Kantian' in the previous 
section--- seems to be that we are Boolean beings, 
and that to describe and communicate the results of experiments 
we need to specify a particular Boolean 
perspective, which we do via the specification of a classically 
describable macroscopic experimental arrangement. 
 
I propose that what we ought to aim for in interpreting quantum mechanics as a 
principle theory is an interpretation that preserves as much as 
we can of our realist 
intuitions about possibility, actuality, and probability, 
subject to the constraints of the `no go' hidden 
variable theorems, which limit the applicability of these intuitions 
in a quantum world. As it turns out, the Copenhagen interpretation is 
one of a well-defined range of 
interpretations of this sort.

We know from the `no go' theorems that 
we cannot generate the 
probabilities defined by a quantum state, for ranges of values of the 
observables of a quantum system, from a measure function on 
a probability space of elements representing
all possible assignments of values to these observables, if the value 
assignments are required to satisfy certain locality or 
structure-preserving constraints. What this 
means is that, if we accept the constraints as reasonable and require 
that \emph{all} observables are assigned values, we cannot 
interpret the quantum probabilities as measures of ignorance of the 
actual unknown values of these observables. In fact, the `no go' 
theorems show that there are no 
consistent value assignments at all to certain well-chosen finite sets 
of observables, quite apart from the question of generating the 
quantum probabilities as measures over possible value 
assignments. Note that the `irreducibility' of quantum 
probabilities  in this sense arises from certain structural features 
of Hilbert space, brought out for the first time by the `no go' 
theorems. It does not follow from earlier considerations, such as 
Heisenberg's uncertainty principle, which refers only to a reciprocal 
relationship between the statistical distributions of certain 
observables for a given quantum state, and says nothing about 
hypothetical value assignments to observables. 

We also know that if 
we consider any quantum state $|\psi\rangle$  and any single observable $R$, 
the probabilities defined by $|\psi\rangle$ for ranges of values of $R$ 
\emph{can} be represented in this way, essentially because the
Hilbert space subspaces 
associated with the ranges of values of a single observable generate a 
Boolean algebra (or Boolean lattice). 
So the possibility of consistent value assignments, 
or the representation of quantum probabilities as measures over such 
value assignments, must fail somewhere between considering a single observable 
and all observables. The relevant question to ask is therefore this:  
beginning with an arbitrary quantum 
state $|\psi\rangle$ and the Boolean lattice generated by a single 
observable $R$, how large a set of observables can we add to $R$ 
before things `go wrong,' that is, before we run up against the `no go' theorems? 
More precisely, a 2-valued 
homomorphism on the Boolean sublattice generated by $R$ is a map 
that assigns 1's and 0's to the elements of the sublattice in 
structure-preserving way, and so defines an assignment of values to 
the observables associated with the sublattice. 
What is the maximal lattice extension 
${\cal D}(|\psi\rangle, R)$ of this Boolean lattice, 
generated by the subspaces associated 
with ranges of values of observables, on which we can represent the 
probabilities defined by $|\psi\rangle$, for the ranges of values of $R$ and 
these additional observables, in terms of a measure over 2-valued 
homomorphisms on ${\cal D}(|\psi\rangle, R)$? 

This question is answered by a uniqueness theorem first proved by Bub 
and Clifton (1996). The theorem provides an answer to the question, on the 
further assumption that ${\cal D}(|\psi\rangle, R)$ is to be invariant under 
automorphisms of the lattice ${\cal L}$ of all subspaces of 
Hilbert space that 
preserve the ray representing the state $|\psi\rangle$ and the 
`preferred observable' $R$. The original proof (reproduced in 
Bub (1997)) involved a `weak separability' assumption 
(introduced to avoid a 
dimensionality restriction) that required several preliminary 
definitions and considerably complicated the formulation of the 
theorem. Sheldon Goldstein has pointed out 
that the proof goes through without this assumption. For details, see 
the revised 
proof by Bub, Clifton, and Goldstein following this 
article. This analysis, in terms of the lattice structure of 
finite-dimensional 
Hilbert spaces, has now been generalized by Clifton (1999) 
and Halvorson and 
Clifton (1999) to 
cover continuous observables and mixed states in the 
general framework of C*-algebras.

It turns out that the sublattice 
${\cal D}(|\psi\rangle, R) \subset {\cal L}$ 
is unique. It is the sublattice generated by (i)the non-zero projections 
of $|\psi\rangle$ onto the $R$-eigenspaces, and (ii) all the rays in the 
subspace orthogonal to the span of these projections, by lattice 
completion. (In the general case, Halvorson 
and Clifton (1999) show that uniqueness fails for certain mixed 
states and certain choices of $R$.) 
In fact, ${\cal D}(|\psi\rangle, R)$
is a generalization of the 
orthodox (Dirac-von Neumann) 
sublattice obtained by taking all the subspaces assigned 
probability 1 or 0 by $|\psi\rangle$ as representing determinate
properties of the system in the state $|\psi\rangle$, and all 
other properties as indeterminate (so that the propositions asserting 
that the system possesses these properties in the state $|\psi\rangle$ are 
neither true nor false or, as a physicist might say, `meaningless' 
in the state $|\psi\rangle$). 

From the standpoint of the theorem, the 
Dirac-von Neumann sublattice is obtained by choosing $R$ as the 
identity observable 
$I$, but this choice leads to the measurement 
problem, as I show in Bub (1997, sections 4.1 and 5.1). 
Other choices for $R$ can be associated with various 
non-orthodox `no collapse' interpretations of quantum mechanics, 
for example Bohm's hidden 
variable theory and modal interpretations that exploit the biorthogonal 
decomposition theorem (cf. Bub (1997, Chapter 6)).

The choice of some preferred observable $R$ other than $I$ requires 
introducing a dynamics for the evolution of \emph{actual properties} or 
\emph{actual values} of 
observables associated with the `determinate sublattice' 
${\cal D}(|\psi\rangle,R)$, as this sublattice evolves over time with the 
unitary evolution of $|\psi\rangle$ as a solution to Schr\"{o}dinger's 
equation of motion. Of course, this dynamics for actual values will 
have to mesh with the Schr\"{o}dinger dynamics tracked by $|\psi\rangle$. I 
sketch such a dynamics in Bub (1997, Chapter 5). It turns out to be a stochastic 
dynamics that reduces to the deterministic dynamics Bohm 
introduced for the actual 
values of position in configuration space in his 1952 hidden variable 
theory, if we take $R$ as continuous position in 
configuration space. The issue of a modal dynamics has been 
investigated in full generality by Bacciagaluppi and Dickson 
(1999).

What determines the choice of $R$ if $R \neq I$? An interpretation of quantum 
mechanics requires the identification of a 
\emph{suitable} preferred observable that we can take as determinate 
and in terms of which we can interpret certain physical processes as 
measurements yielding distributions of determinate pointer readings. 
I propose that such a preferred observable is picked out by the 
phenomenon of environmental decoherence. It is this phenomenon that 
guarantees the possibility of measurement: 
dynamical change of a certain sort that characterizes the transfer of 
information between systems. Recent discussions in the 
literature by Zurek (1991, 1993) and others develop sophisticated models of 
the interaction between the sorts of systems we use as a measuring 
instruments and the typical environment in which such instruments are 
used. The essential feature of these models is that the 
interaction that takes place in our world, between a system we 
are able to use as a measuring instrument and the environment, 
selects a particular observable as a suitable measurement 
pointer. This is an observable of the measuring instrument 
that becomes correlated with an
observable of the measured system, for correlations that remain stable over 
time while the measuring instrument undergoes 
an effective `monitoring' by the environment. Such an instrument observable must 
commute with the instrument-environment interaction Hamiltonian, and 
because of the nature of typical instrument-environment interactions 
in our universe it appears that a coarse-grained position-type 
observable is selected. So the fact that `the 
environment acts, in effect, as an observer continuously monitoring 
certain preferred observables which are selected mainly by the 
system-environment interaction hamiltonians' (Zurek, 1993, p. 290) 
is a contingent dynamical feature of our quantum world that
guarantees the existence of a preferred determinate observable, and 
hence the possibility of measurement and the exchange of information.

In terms of the uniqueness theorem, the Copenhagen interpretation can 
be understood as involving the claim that we can only describe and 
communicate the results of experiments in our non-Boolean quantum 
world by specifying a particular Boolean perspective, associated with 
a classically describable experimental arrangement that, in effect, 
selects a particular preferred observable $R$ as determinate. In the 
context of the experimental arrangement, we are entitled to speak of 
\emph{this} observable as having a determinate value, but the values 
of complementary observables, associated with incompatible 
experimental arrangements, are indeterminate, constrained only by the 
uncertainty principle. What is right about this view is that 
description and communication in a quantum world requires an 
$R$-perspective, that is, a determinate sublattice 
${\cal D}(|\psi\rangle,R)$, 
but this sublattice need not be a Boolean algebra. Moreover, this 
sublattice need not be stipulated in terms of the resources of 
classical mechanics, and a conventional `cut' between the observed 
part of the universe and the observer, 
but can be generated from processes internal to quantum mechanics on 
the basis of the dynamics alone, without requiring any privileged status 
for observers.

The significance of interpreting quantum mechanics as a principle 
theory can now be understood along the following lines: The move from 
special relativity as a principle theory of geometric structure to 
general relativity involves the insight that geometry is not only 
empirical but \emph{dynamical}. That is, the transition from 
classical mechanics to special relativity, as the theoretical framework 
incorporating Einstein's special principle of relativity and the 
principle of the constancy of the velocity of light, requires that 
geometry cannot be \emph{a priori}. It makes sense to ask: what is the 
geometry of the world? Extending special relativity 
to general relativity leads to a relativistic theory of gravitation 
in which the geometry of our universe 
is not a fixed Euclidean geometry, as we supposed classically, 
but rather a non-Euclidean 
geometry that changes dynamically as the distribution 
of mass in the universe changes.  The import of the 
uniqueness theorem is that, just as the transition from classical 
mechanics to relativistic mechanics as a principle theory of 
geometric structure leads to 
the conclusion that
geometry is dynamical, so the transition from classical mechanics to 
quantum mechanics as a principle theory of logical or possibility structure 
leads to the conclusion 
that possibility is 
dynamical: the possibility structure of our universe is not a fixed, 
Boolean structure, as we supposed classically, but is in fact a 
non-Boolean structure that changes dynamically. The unitary 
Schr\"{o}dinger evolution of the quantum state in time tracks the 
evolution of this possibility structure as a dynamically changing sublattice 
${\cal D}(|\psi\rangle,R)$ in 
the lattice of all subspaces of Hilbert space. So the Schr\"{o}dinger 
time-dependent equation characterizes the temporal evolution of what 
is \emph{possible}, not what is \emph{actual} at time $t$. 

In a classical world, change is 
described by the evolution over time of what is actual, 
where what is 
actually the case at time $t$ is selected by a 2-valued 
homomorphism---the classical state---as a temporally evolving 
substructure against the background of a \emph{fixed} Boolean lattice of 
possibilities. In a quantum world, what is actually the 
case at time $t$ is selected by a 2-valued homomorphism as a 
temporally evolving substructure 
on a \emph{dynamically changing} background of possibilities. So in a quantum 
world there is a dual 
dynamics: the Schr\"{o}dinger dynamics for the evolution of 
possibility, and a dynamics for how what is actually the case changes 
with time, which must mesh with the evolving possibility structure and 
turns out to be a generalization of Bohmian dynamics. 
I shall refer to the state in the sense of a catalogue of properties 
selected by a 2-valued homomorphism on ${\cal D}(|\psi\rangle, R)$
as the `property state' to distinguish this notion of state from 
the quantum state $|\psi\rangle$.

\section{Entanglement}

In terms of the view outlined in the previous section, we can now 
understand `entanglement'---what 
Schr\"{o}d\-inger (1935, p. 555)
called `\emph{the} 
characteristic trait of quantum mechanics, the one that enforces its 
entire departure from classical lines of thought'---as 
arising from the dynamical evolution of the possibility 
structure of composite systems, 
through the tensor product representation of their quantum states. 
(Note that entangled states do not 
occur in a classical wave theory, where the states of composite systems 
are Cartesian products of the subsystem states.)

I begin by considering the significance of Bohr's reply to the 
Einstein-Podolsky-Rosen argument (1935) from this standpoint, 
applied to the example of two separated spin-$\frac{1}{2}$ systems in 
the singlet spin state. The essential point to understand here is the 
peculiar quantum mechanical nonlocality, or better, nonseparability 
(in Bohr's terminology, `wholeness') of the separated systems in the 
`entangled' singlet state. I show how this phenomenon arises as a 
feature of the dynamical evolution of the possibility structure, from 
the perspective of the determinate sublattice selected by a 
coarse-grained preferred position observable, following the discussion in the 
previous section concerning the selection of the preferred observable $R$. 
I go on to consider the recent application of 
entangled states to quantum teleportation, in which a shared entangled 
state between two parties, Alice and Bob, allows the instantaneous 
transfer of a quantum state from Alice to Bob, with no violation of 
relativistic principles. (See Lo, Popescu, and 
Spiller (1998) for an account and references to the original 
experiments.) The puzzle here is how the information gets from Alice 
to Bob. I show how this puzzle is resolved in a similar way 
by considering the effect 
of Alice's operations (as unitary transformations) on the determinate 
sublattice. 

Consider two separated spin-$\frac{1}{2}$ particles, $S_{1}$ and 
$S_{2}$, in the singlet spin state. To bring out the conceptual 
issue clearly suppose, for simplicity, that the 
preferred observable $R$ is a discrete, coarse-grained 
position observable with eigenvalues corresponding to the pairs of values of 
$R_{1}$ and $R_{2}$, the two discrete position observables of $S_{1}$ and 
$S_{2}$. Then the quantum state of 
$S_{1}+S_{2}$ can be represented schematically as:
\begin{equation}
|\Psi_{0}\rangle = (\frac{1}{\sqrt{2}}|+\rangle_{1}|-\rangle_{2} - 
\frac{1}{\sqrt{2}}|-\rangle_{1}|+\rangle_{2})|r_{0}\rangle_{1}|r_{0}\rangle_{2},
\end{equation}
where $|+\rangle$ and $|-\rangle$ denote spin component eigenstates 
in the $z$-direction, and $|r_{0}\rangle_{1}$ and $|r_{0}\rangle_{2}$ 
represent the initial `zero' positions of the two particles. Now 
suppose that a unitary transformation is applied at $S_{1}$ that 
entangles eigenstates of $R_{1}$ with spin component eigenstates in 
the $z$-direction, corresponding to what is sometimes called 
a `premeasurement' (without 
`collapse') of $z$-spin on particle $S_{1}$. This yields the 
transition:
\begin{equation}
|\Psi_{0}\rangle  \stackrel{U\otimes I}{\longrightarrow}  |\Psi\rangle 
= 
 \frac{1}{\sqrt{2}}|+\rangle_{1}|-\rangle_{2}|r_{+}\rangle_{1}|r_{0}\rangle_{2}
  - 
\frac{1}{\sqrt{2}}|-\rangle_{1}|+\rangle_{2})|r_{-}\rangle_{1}|r_{0}\rangle_{2},
\end{equation}
where $U$ is defined on the tensor product of the spin Hilbert 
space of $S_{1}$ and the position Hilbert space of $S_{1}$, and $I$ 
is the identity on the corresponding product space of $S_{2}$.

The eigenspaces of the preferred observable $R$ are the subspaces in 
the tensor product Hilbert space of $S_{1}+S_{2}$ that 
correspond to particular values for $R_{1}$ and 
$R_{2}$. For simplicity, suppose that $R_{1}$ and $R_{2}$ can each 
take one of three possible values, $-$, $0$ , or $+$, so that there are
nine eigenspaces for $R$. While 
${\cal D}(|\Psi_{0}\rangle,R)$ does not contain $z$-spin properties for 
$S_{1}$ or $S_{2}$ (because the non-zero projections of 
$|\Psi_{0}\rangle$ onto the $R$-eigenspaces yield only the property 
corresponding to the superposition 
$\frac{1}{\sqrt{2}}|+\rangle_{1}|-\rangle_{2} - 
\frac{1}{\sqrt{2}}|-\rangle_{1}|+\rangle_{2}$), 
the sublattice ${\cal D}(|\Psi\rangle,R)$ does (because the non-zero 
projections of $|\Psi\rangle$ onto the $R$-eigenspaces yield the 
properties corresponding to $z$-spin product states $|+\rangle_{1}|-\rangle_{2}$ 
and $|-\rangle_{1}|+\rangle_{2}$). So 
a unitary 
transformation at $S_{1}$, which applies also to the composite 
entangled 
system $S_{1}+S_{2}$, can make a property at $S_{2}$ determinate that 
was not determinate before, via the dynamical 
evolution of possibilities open to the system $S_{1}+S_{2}$. There is 
no violation of the `no signalling' requirement of special relativity, 
because the physical process at $S_{1}$ does not change any 
determinate or actual spin-component 
value of $S_{2}$: the determinate sublattice ${\cal D}(|\Psi_{0}\rangle,R)$ 
does not contain spin-component properties of $S_{2}$ (or of $S_{1}$). 
Rather, the unitary transformation at $S_{1}$ results in the dynamical 
evolution of the determinate sublattice to a determinate sublattice 
containing spin-component properties. As Bohr (1935, p. 699) put it in 
his reply to 
Einstein, Podolsky, and Rosen:
\begin{quote}
Of course there is in a case like that just considered no question of 
a mechanical disturbance of the system under investigation during the 
last critical stage of the measuring procedure. But even at 
this stage there is essentially the question of \emph{an influence on 
the very conditions which define the possible types of predictions 
regarding the future behaviour of the system.}
\end{quote}

If we accept that a quantum world differs from a classical world 
in just this way---that a reversible local unitary transformation can change 
what is possible, but not what is actually the case, nonlocally---
then we can understand how the 
phenomenon of quantum teleportation is possible. 

In a teleportation 
protocol, 
Alice and Bob initially share a singlet state:
\begin{equation}
(\frac{1}{\sqrt{2}}|+\rangle_{A}|-\rangle_{B} - 
\frac{1}{\sqrt{2}}|-\rangle_{A}|+\rangle_{B})|r_{0}\rangle_{A}|r_{0}\rangle_{B},
\end{equation}
where $|r_{0}\rangle_{A}$ and $|r_{0}\rangle_{B}$ 
represent the initial `zero' positions of Alice's particle $A$ and Bob's 
particle $B$. Alice 
is required to teleport a particle $C$ in an arbitrary quantum state 
\begin{equation}
|\psi\rangle_{C} = c_{+}|+\rangle_{C} + c_{-}|-\rangle_{C}
\label{eq:psi-teleport}
\end{equation}
to Bob. In effect, she is required to 
`fax' information to Bob that will enable him to reconstruct the 
quantum state of $C$ from the raw material he has on hand: his half 
of the shared entangled state. Remarkably, Alice can do this by 
transferring just two bits of classical information to Bob, vastly 
less than the amount of information required to fully specify the quantum 
state of $C$. The puzzle is: how does the rest of the information get 
from Alice to Bob? (See, for example, Jozsa's discussion in Lo, 
Popescu, and Spiller (1998).)

The initial quantum state of the 
system $A+B+C$ is:
\begin{equation}
|\Phi_{0}\rangle = |\psi\rangle_{C}(\frac{1}{\sqrt{2}}|+\rangle_{A}|-\rangle_{B} 
- 
\frac{1}{\sqrt{2}}|-\rangle_{A}|+\rangle_{B})|r_{0}\rangle_{A}|r_{0}\rangle_{B}%
\label{eq:Phi-initial}
\end{equation}
This can be expressed as:
\begin{eqnarray}
|\Phi_{0}\rangle & = & \frac{1}{2}|1\rangle_{AC}(c_{+}|+\rangle_{B} + 
c_{-}|-\rangle_{B}) |r_{0}\rangle_{A}|r_{0}\rangle_{B}\nonumber \\
 & &  + \frac{1}{2}|2\rangle_{AC}(-c_{+}|+\rangle_{B} + 
c_{-}|-\rangle_{B})|r_{0}\rangle_{A}|r_{0}\rangle_{B} \nonumber \\
& & + \frac{1}{2}|3\rangle_{AC}(c_{-}|+\rangle_{B} + 
c_{+}|-\rangle_{B})|r_{0}\rangle_{A}|r_{0}\rangle_{B} \nonumber\\
& & + \frac{1}{2}|4\rangle_{AC}(-c_{-}|+\rangle_{B} + 
c_{+}|-\rangle_{B})|r_{0}\rangle_{A}|r_{0}\rangle_{B},
\end{eqnarray}
where $|1\rangle, |2\rangle, |3\rangle, |4\rangle$ are the Bell states:
\begin{eqnarray}
|1\rangle & = & \frac{1}{\sqrt{2}}(|+\rangle|-\rangle - 
|-\rangle|+\rangle) \\
|2\rangle & = & \frac{1}{\sqrt{2}}(|+\rangle|-\rangle + 
|-\rangle|+\rangle) \\
|3\rangle & = & \frac{1}{\sqrt{2}}(|+\rangle|+\rangle - 
|-\rangle|-\rangle) \\
|4\rangle & = & \frac{1}{\sqrt{2}}(|+\rangle|+\rangle + 
|-\rangle|-\rangle) 
\end{eqnarray}

The Bell states form an orthonormal basis in the Hilbert space 
${\cal H}_{AC}$. Suppose Alice applies a unitary transformation in 
 ${\cal H}_{AC}$ that entangles the position of A with the Bell states. 
That is, she applies a unitary transformation that corresponds to a 
premeasurement (with the position of $A$ as `pointer') 
of an observable $Q$ of $A+C$ that has the Bell states as eigenstates.
This results in the transition:
\begin{eqnarray}
|\Phi_{0}\rangle  \rightarrow  |\Phi\rangle & =  
& \frac{1}{2}|1\rangle_{AC}|r_{1}\rangle_{A} (-c_{+}|+\rangle_{B} - 
c_{-}|-\rangle_{B}) |r_{0}\rangle_{B}\nonumber \\
 &  &  + \frac{1}{2}|2\rangle_{AC}|r_{2}\rangle_{A}(-c_{+}|+\rangle_{B} + 
c_{-}|-\rangle_{B})|r_{0}\rangle_{B} \nonumber \\
 &  & + \frac{1}{2}|3\rangle_{AC}|r_{3}\rangle_{A}(c_{-}|+\rangle_{B} + 
c_{+}|-\rangle_{B})|r_{0}\rangle_{B} \nonumber\\
 &  & + \frac{1}{2}|4\rangle_{AC}|r_{4}\rangle_{A}(-c_{-}|+\rangle_{B} + 
c_{+}|-\rangle_{B})|r_{0}\rangle_{B}\label{1}
\end{eqnarray}

Consider now the determinate sublattice ${\cal D}(|\Phi\rangle,R)$. 
To simplify the analysis here, suppose that each of four possible 
pointer positions, 
that is, $A$-positions $|r_{1}\rangle_{A}, \cdots, |r_{4}\rangle_{A}$, 
is associated with a 2-valued 
homomorphism on the sublattice, that is, a distinct property state:
an assignment of truth 
values to the propositions in the sublattice that selects a 
catalogue of the actual properties of the system $A+B+C$. 
When Alice ascertains the position of the pointer, she 
obtains two bits of information about the position (which, as the 
result of the premeasurement unitary transformation, has one of four 
possible positions, each with equal probability), and 
hence two bits of information about the correlated property state.

From these two bits of information (conveyed to Bob by Alice via a 
classical channel), Bob can select one of four possible unitary 
transformations on the Hilbert space ${\cal H}_{B}$, with the 
properties:
\begin{eqnarray}
-c_{+}|+\rangle_{B} - c_{-}|-\rangle_{B} & 
\stackrel{U_{B(1)}}{\rightarrow} & |\psi\rangle_{B} \\
-c_{+}|+\rangle_{B} + c_{-}|-\rangle_{B} &
\stackrel{U_{B(2)}}{\rightarrow} & |\psi\rangle_{B} \\
c_{-}|+\rangle_{B} + c_{+}|-\rangle_{B} &
\stackrel{U_{B(3)}}{\rightarrow} & |\psi\rangle_{B} \\
-c_{-}|+\rangle_{B} + c_{+}|-\rangle_{B} &
\stackrel{U_{B(4)}}{\rightarrow} & |\psi\rangle_{B}
\end{eqnarray}
where $|\psi \rangle_{B}$ is given by (\ref{eq:psi-teleport}).
Since $|\Phi\rangle$ is a product state on ${\cal H}_{AC}\otimes 
{\cal H}_{B}$, the effect of applying the unitary transformation 
$U_{B(i)}, i = 1, \ldots, 4$, on ${\cal H}_{B}$, which is equivalent to 
$U_{i} = I_{AC}\otimes U_{B(i)}$ on ${\cal H}_{AC}\otimes 
{\cal H}_{B}$, 
is to transform 
$|\Phi\rangle$ to $U_{(i)}|\Phi\rangle = |\Phi_{i}\rangle$. 
The property state on 
${\cal D}(|\Phi_{i}\rangle,R)$, for the value of $i$ corresponding to the 
actual pointer position, is generated by a 2-valued 
homomorphism that assigns 1 to the component 
\[
|i\rangle_{AC}|r_{i}\rangle_{A}|\psi\rangle_{B}|r_{0}\rangle_{B}
\]
of $|\Phi_{i}\rangle$.
That is, the effect of the transformation is to instantiate the 
property represented by the projection operator 
$|\psi\rangle\langle\psi|$ on ${\cal H}_{B}$. So Bob requires only 
two bits of information to reconstruct the property state on $A+B+C$ 
that contains the $B$-property represented by the 1-dimensional 
subspace spanned by $|\psi\rangle_{B}$ in ${\cal H}_{B}$. The 
puzzle about how this information suffices for Bob to 
reconstruct the teleported state is resolved 
once we see that Alice's local unitary transformation 
alters the global determinate 
sublattice to one in which there are four property states, with equal 
probability, related in known ways to the required property state.

\section{Decoherence}

On the Dirac-von Neumann interpretation, 
Alice's measurement is supposed to `collapse' the (global) state onto one of the 
branches in the superposition $|\Phi\rangle$. But this collapse is surely 
an unacceptably \emph{ad hoc} modification of the quantum dynamics, 
introduced solely to pick out the branch of the 
superposition that accords with the properties that \emph{actually} 
obtain, according to the Dirac-von Neumann prescription. And there is 
no teleportation on this interpretation without the collapse.

A currently 
fashionable view is to appeal to environmental decoherence. 
If the Hamiltonian characterizing 
the interaction with the environment commutes with $R$, each 
of the pointer states $|r_{i}\rangle$, for $i = 1, 2, 3, 4,$ in 
$|\Phi\rangle$, becomes coupled with a state of the environment, where these 
environmental states very rapidly approach orthogonality. To exhibit 
interference between the different branches of the superposition, one 
would have to perform an appropriate experiment on the system together 
with the environment, which is for all practical purposes impossible. 
On this basis, it is argued that because the different 
branches of the superposition effectively no longer interfere after 
the entanglement of the pointer position with the eigenstates of the 
measured observable $Q$, we are entitled to take one of the branches 
as the actual branch. But nothing in the Dirac-von Neumann interpretation 
sanctions this move---nothing distinguishes one of the branches as 
privileged in this way. 
Moreover, it follows from the uniqueness theorem that it is 
 \emph{inconsistent} to take any of the properties associated 
with one of the branches of the superposition as actual on the basis of the 
Dirac-von Neumann interpretation, together with 
the other properties that are taken as actual on the basis of this 
interpretation (that is, the properties assigned 
unit probability by the total quantum state). A further move is to 
take all the branches as equally `actual' in some sense, but the notion of 
actuality, as distinct from possibility, then becomes empty. 

On the view I have sketched above, decoherence means that the 
temporal evolution of the property state will, for all practical 
purposes, be independent of those parts 
of the total quantum state $|\Phi\rangle$ in ${\cal H}_{A+B+C+E}$ 
that are effectively orthogonal to 
the component corresponding to the actual pointer 
position. For a discussion, see Bub (1997, section 5.4). So when 
Alice obtains her two bits of information about the pointer position 
and conveys this information to Bob, Bob knows that the 
quantum state of $B$ is effectively one of four possible states, 
a factor state of the 
component of the total quantum state of the composite system $A+B+C+E$ that 
corresponds to Alice's pointer position. There is an effective 
collapse of the state as a result of the interaction of the pointer 
system with the environment, and it is because of this effective collapse that 
one can say that, for all practical purposes, the quantum state of $C$ 
has been transferred to $B$: the evolution of the property state of 
$B$ will be determined, effectively, by this quantum state. (Note that 
\emph{this} decoherence argument is not available on the 
Dirac-von Neumann version 
of the orthodox interpretation.) 

The decoherence account of 
measurement is supposed to validate the 
projection postulate or the collapse of the quantum state in a 
measurement process. Now, a 
measurement interaction between a system $S$ 
and a measuring instrument $M$, followed by a virtually instantaneous 
interaction between $M$ and the environment $E$ (via a Hamiltonian 
that commutes with the pointer observable), yields a quantum state 
for $S+M+E$ that, in virtue of the nature of the interaction between 
$M$ and $E$, takes a certain form. Expressing the state as a density 
operator and tracing over ${\cal H}_{E}$, yields a reduced density 
matrix $W_{S+M}$ for $S+M$ that is effectively diagonal in the 
pointer basis: the off-diagonal elements decay almost instantaneously 
to zero. But the fact that the density operator $W_{S+M}$, obtained by 
`ignoring' or `averaging over' the environment takes the form of a 
mixture with respect to properties associated with the pointer basis 
not only fails to account for the occurrence of \emph{just one} of 
these events, but is actually \emph{inconsistent} with such an 
occurrence. If we consider the origin of the mixture, and the 
Dirac-von Neumann rule for relating the quantum state to a property 
state, the property state defined by the quantum state 
$|\Phi\rangle$ 
of $S+M+E$ selects a determinate sublattice 
${\cal D}(|\Phi\rangle,I)$ in ${\cal H}_{S+M+E}$. This determinate 
sublattice is \emph{maximal} by the uniqueness theorem, and so we 
cannot add to it properties that are determinate, via the Dirac-von 
Neumann rule, on the basis of one of the states in the mixture. One 
can, in fact, show quite easily, independently of the theorem, that if 
we add \emph{any} proposition, represented by a subspace in 
${\cal H}_{S+M+E}$, to the determinate sublattice 
${\cal D}(|\Phi\rangle,I)$, then we have to add the propositions represented 
by \emph{every} subspace in ${\cal H}_{S+M+E}$ to 
${\cal D}(|\Phi\rangle,I)$, and this would, of course, generate a 
Kochen-Specker contradiction.

The appeal of the decoherence solution to the measurement problem 
derives from the belief that we can interpret the reduced density 
operator $W_{S+M}$ as representing the occurrence of a particular 
event---the event associated with a particular pointer reading and the 
instantiation of the correlated properties of $S$---with the terms 
along the diagonal of the density matrix in the position basis 
representing a measure of our ignorance as to the actual event. The 
suggestion is that the procedure of tracing over the environment is 
analogous to the procedure of deriving a probability of $\frac{1}{2}$ 
for `heads' and $\frac{1}{2}$ for `tails' in a coin toss experiment by 
averaging over the uncontrolled and unmeasured degrees of freedom of 
the environment of the coin. Zurek (1996, p. 39), 
for example, refers to the 
procedure by which one derives the reduced density operator as  `ignoring 
(tracing over) the uncontrolled (and unmeasured) degrees of freedom 
of the environment.' But the two procedures are not at all analogous. 
When we `ignore' the environment  to claim that the probability of 
getting `heads' on a particular toss of the coin is $\frac{1}{2}$, we 
can also claim that we \emph{do} in fact get \emph{either} `heads' 
\emph{or} `tails' on each particular toss, and whether we get `heads' 
or `tails' on a particular toss depends on the precise values of 
certain environmental parameters, which we do not attempt to control 
or measure. But in the quantum mechanical case, we cannot claim that 
taking full account of the environment on each particular occasion 
would fix the value of the pointer as one particular value. Taking 
full account of the environment will, of course, give us back the 
pure state of $S+M+E$ from which the mixture $W_{S+M}$ was derived. 
And this state is inconsistent with the occurrence of 
events associated with definite pointer readings on the orthodox 
interpretation.

\section{Instrumentalism}

In the previous sections, I argued for an interpretation of 
quantum mechanics as a principle theory, and I endorsed the Copenhagen 
interpretation as implicitly taking a similar view. Of course, the 
Copenhagen interpretation is more commonly given an instrumentalist 
reading. Here I want to 
take issue with instrumentalist solutions to the interpretative 
problems of quantum mechanics.

Many physicists, for example, Peres (1980, 1986, 1988, 1993, 1998) and van 
Kampen (1988), reject the Dirac-von Neumann version of the 
orthodox interpretation and profess to champion the Copenhagen 
interpretation in Bohr's formulation. Generally, Bohr's sometimes 
obscure pronouncements are given an instrumentalist slant. In a 
review of my book, Peres remarks (1998, pp. 612--613):

\begin{quote}
The tacit assumption made by Bub (as well as by many authors who tried 
to come to grips with [the interpretation] problem) is that the wave 
function is a genuine physical entity, not just an intellectual tool 
invented for the purpose of computing probabilities. \ldots In the 
theoretical laboratory, wave functions are routinely employed by 
physicists as mathematical tools, which are useful for predicting 
probabilities for the various possible outcomes of a measurement 
process.
\end{quote}

Both Peres and van Kampen develop similar accounts of measurement in 
quantum mechanics, which they see as consistent with Bohr's position. 
A measurement apparatus is treated as a macroscopic system with many degrees of 
freedom, prepared in a certain macrostate. As van Kampen emphasizes 
(1988, p. 101):
\begin{quote}
When a macroscopic pointer indicates a macroscopic point on a dial 
the number of microscopic eigenstates involved has been estimated by 
Bohm (1951, Chapter 4) to be $10^{50}$. 
When the observer shines in light in 
order to read the position of the pointer, the photons do perturb the 
$\psi$ of the pointer, but the perturbation does not affect the 
macrostate. The vector $\psi$ is moved around a bit in these $10^{50}$ 
dimensions but its components outside the subspace remain negligible. 
That is the reason why macroscopic observations can be recorded 
objectively, independently of the observations and the observer, and 
may therefore be the object of scientific study. The lilliputian 
measurements of Heisenberg (1949, Chapter II).
and von Neumann do not apply to 
experiments with macroscopic systems.
\end{quote}

The `lilliputian measurements of Heisenberg and von Neumann' involve 
unitary transformations of the quantum state. But then, as Peres 
notes (1986, p. 691), `nothing happens':
\begin{quote}
The two electrons in the ground state of the helium atom are 
correlated, but no one would say that each electron `measures' its 
partner. In general, if we have a piece of hardware which can be used 
as a measuring apparatus, we must choose one of the following 
alternatives: Either let it work in a noisy environment (including 
its own internal `irrelevant' degrees of freedom) or let it be 
perfectly prepared and isolated, and described by the Schr\"{o}dinger 
equation. In the latter case, that piece of hardware loses its status 
of `measuring apparatus.' This is just a matter of having consistent 
definitions: A measuring apparatus \emph{must} have macroscopically 
distinguishable states, and the word `macroscopic' has just been 
defined as `incapable of being isolated from the environment.'

Yet anyone is free to imagine a perfect world, completely and 
exhaustively described \ldots. In that world, there is neither noise 
nor irreversibility\ldots. In that perfect world, nothing happens 
and, in particular, there are no measurements.
\end{quote}

For Peres (1986, p. 691), the collapse of the quantum state 
on measurement `is not a 
physical process, but simply the acquisition of fresh knowledge about 
a physical system. It is a change of our description, whereby we 
\emph{return from a Gibbs ensemble to a single object.'}

Knowledge of what? A procedure that is quite unobjectionable 
classically, where there is a fact of the matter about which we are 
initially ignorant and come to know via a measurement, is quite 
incomprehensible in a purely quantum description, where `nothing happens.' The 
view is tenable only if we smuggle in some notion of determinateness 
or actuality. But what is the principle involved here? We can't simply 
assert, by fiat, that macrostates are determinate, that something or 
other happens at the macrolevel but not at the microlevel. That would 
be tantamount to saying that whether or not something \emph{happens}, 
or whether or not something is actually the case, depends on whether 
or not we \emph{ignore} certain aspects of the world as `noise.' So 
whether or not an event takes place would not be an objective feature 
of the world but would depend entirely on features of our description 
of phenomena---features that have to do with \emph{our} limited 
technological and mathematical abilities, and \emph{our} interests.

Van Kampen sees the collapse a little differently. He considers a 
measuring apparatus that can detect whether an electron has passed 
through a region $U$ in space. The apparatus consists of an atom in an 
excited state, together with an electromagnetic field. The electron 
distorts the state allowing the emission of a photon, which can be 
detected by catching it on a photographic plate. Van Kampen shows 
that there is a term in the quantum state of the total system 
representing a wave emanating from the region $U$, associated with the 
triggering of the measurement apparatus by the electron. He 
concludes (1988, p. 106):
\begin{quote}
\emph{This is the collapse of the wave function:} when the apparatus 
has observed the electron to be in $U$ the electron wave function is 
no longer the initial $\phi$ but is replaced by $\psi_{k}$. Thus the 
collapse is not an additional postulate and has nothing to do with a 
change of my knowledge or some such anthropomorphic consideration.
\end{quote}

But what does it mean for the apparatus to `observe the electron to be in 
$U$'? In a purely quantum description there is nothing that 
selects this apparatus event as privileged. And why should an `observation' in 
this sense require the quantum state to be replaced by a component of 
the total state, unless the Dirac-von Neumann interpretation is 
invoked implicitly? 

Of course, a purely instrumental interpretation of quantum 
mechanics---or any theory---is a 
consistent view. But then, as Einstein remarked in a letter to 
Schr\"{o}dinger (see Przibram (1967, p. 39):
\begin{quote}
If that were so then physics could only claim the interest of 
shopkeepers and engineers; the whole thing would be a wretched bungle.
\end{quote}

\bigskip

\noindent
\emph{Acknowledgement}---The present version of the paper reflects 
helpful critical comments by two anonymous referees and especially the editor, 
Jeremy Butterfield.

\section*{References}

\begin{list}{}{\itemindent -.3in}

\item Bacciagaluppi, G. and Dickson, M. (1999) `Dynamics for Density 
Operator Interpretations of Quantum Mechanics,' quant-ph/9711048; 
`Dynamics for Modal Interpretations,' \emph{Foundations of Physics}, 
forthcoming, 1999.

\item Bell, J.S. (1964) `On the Einstein-Podolsky-Rosen Paradox,' 
\emph{Physics} {\bf 1}, 195--200.

\item Bohm, D. (1951) \emph{Quantum Theory} (New York: Prentice-Hall).

\item Bohr, N. (1935) `Can Quantum-Mechanical Description of Physical 
Reality be Considered Complete?' \emph{Physical Review} {\bf 48}, 
696--702.

\item Bohr, N. (1961) \emph{Atomic Physics and Human Knowledge} (New 
York: Science Editions).

\item Bohr, N. (1966) \emph{Essays 1958--1962 on Atomic Physics and 
Human Knowledge} (New York: Vintage Books).

\item Bondi, H. (1964) \emph{Relativity and Common Sense: A New 
Approach to Einstein} (New York: Anchor Books).

\item Bondi, H. (1967) \emph{Assumption and Myth in Physical Theory} 
(Cambridge: Cambridge University Press).

\item Bub, J. (1997) \emph{Interpreting the Quantum World} (Cambridge: 
Cambridge University Press). Revised edition, 1999.

\item Bub, J. and Clifton, R. (1996) `A Uniqueness Theorem for ``No Collapse''
Interpretations of Quantum Mechanics,' \emph{Studies in the History 
and Philosophy of Modern Physics} {\bf 27}, 181--219.

\item Clifton, R. (1999) `Beables in Algebraic Quantum Mechanics,'
in J. Butterfield and C. Pagonis (eds.), \emph{From Physics to
Philosophy}, (Cambridge: Cambridge University
Press).

\item Einstein, A. (1919) `What is the Theory of Relativity,' The London 
\emph{Times}, November 28.

\item Einstein, A. (1954) \emph{Ideas and Opinions} (New York: Bonanza 
Books).

\item Einstein, A., Podolsky, B., and Rosen, N. (1935) `Can 
Quantum-Mechanical Description of Physical Reality be Considered 
Complete?' \emph{Physical Review} {\bf 47}, 777--80.

\item Halvorson, H. and Clifton, R. (1999) `Maximal
Subalgebras of Beables,' quant-ph/9905042.

\item Heisenberg, W. (1949) \emph{The Physical Principles of Quantum 
Theory} (New York: Dover).

\item Kochen, S. and Specker, E.P. (1967) `The Problem of Hidden 
Variables in Quantum Mechanics,' \emph{Journal of Mathematics and 
Mechanics} {\bf 17}, 59--87.

\item Lo, H.-K., Popescu, S.,and Spiller, T. (1998) \emph{Introduction to 
Quantum Computation and Information} (Singapore: World Scientific).

\item M\"{u}ller, F. (1997) `The Equivalence Myth of Quantum 
Mechanics (Parts 1 and 2),' 
\emph{Studies in the History and Philosophy of Modern Physics} {\bf 
28B}, 35--62; 219--249

\item Peres, A. (1980) `Can We Undo Quantum Measurements,' \emph{The 
Physical Review} {\bf D22}, 879--883.

\item Peres, A. (1986) `When is a Quantum Measurement?', \emph{American 
Journal of Physics} {\bf 54}(8), 688--692.

\item Peres, A. (1988) `Schr\"{o}dinger's Immortal Cat,' \emph{Foundations 
of Physics} {\bf 18}, 57--76.

\item Peres, A. (1993) \emph{Quantum Theory: Concepts and Methods} 
(Dordrecht: Kluwer).

\item Peres, A. (1998) Essay Review:\emph{Interpreting the Quantum World}, this 
journal {\bf 29}, 611--620.

\item Przibram, K. (ed.) (1967) \emph{Letters on Wave Mechanics} (New 
York: Philosophical Library).

\item Schr\"{o}dinger, E. (1935) `Discussion of Probability Relations 
Between Separated Systems,' \emph{Proceedings of the Cmabridge 
Philosophical Society} {bf 31}, 555--563.

\item Tomonaga, S.-I. (1968) \emph{Quantum Mechanics Vol. 1: Old 
Quantum Theory} (Amsterdam: North-Holland).

\item Van Kampen, N.G. (1988) `Ten Theorems About Quantum Mechanical 
Measurements,' \emph{Physica A} {\bf 153}, 97--113.

\item Zurek, W. (1991) `Decoherence and the Transition from Quantum 
to Classical,' \emph{Physics Today} {\bf 44}, 36--44.

\item Zurek, W. (1993) `Preferred States, Predictability, 
Classicality, and the Environment-Induced Decoherence,' \emph{Progress 
in Theoretical Physics} {\bf 89}, 281--312.

\end{list}

\end{document}